\documentclass[prb,twocolumn,showpacs,preprintnumbers,amsmath,amssymb]{revtex4}
\usepackage{graphicx}

\begin{document}

\title{Crossed Andreev Reflection \\ in Structures Consisting of a Superconductor with Ferromagnetic Leads}
\author{Taro Yamashita, Saburo Takahashi, and Sadamichi Maekawa}
\affiliation{Institute for Materials Research, Tohoku University, Sendai 980-8577, Japan}

\date{\today}

\begin{abstract}
A theory of crossed Andreev reflection in structures consisting of a superconductor with two ferromagnetic 
leads is presented.  The electric current due to the crossed Andreev reflection strongly depends on the relative 
orientation of the magnetization of two ferromagnetic leads.  It is shown that the dependence of the electric 
current and the magnetoresistance on the distance between two ferromagnetic leads is understood by considering 
the interference between the wave functions in ferromagnets.  The current and the magnetoresistance are 
calculated as functions of the exchange field and the height of the interfacial barriers.  
\end{abstract}

\maketitle

\section{Introduction}
Much attention has been focused on the spin dependent transport through magnetic nanostructures.\cite{maekawaBOOK}  
The tunnel magnetoresistance (TMR) was observed in ferromagnet/ferromagnet (FM/FM) tunnel junctions 
\cite{julliere,maekawa,miyazaki,moodera}.  In ferromagnet/superconductor (FM/SC) tunnel junctions, 
the current flowing thorough the tunnel junctions is spin polarized.\cite{meservey}  
When the spin polarized quasiparticles (QPs) is injected into SC from FM, the superconducting gap is suppressed 
due to the spin accumulation in FM/SC and FM/SC/FM junctions.\cite{vasko,dong,yeh,liu,chen,saburo1,zheng,bozovic}
The detail studies of the spin transport and relaxation in SC have been done.\cite{taro1,saburo2,saburo3}  

In recent years, many theoretical and experimental studies in relation to Andreev reflection \cite{andreev} 
in FM/SC metallic contacts have been done because the spin polarization of conduction electrons is estimated by 
measuring the conductance in this system.
\cite{dejong,kikuchi,hima,zhu,igor,kashiwaya,soulen1,soulen2,ji,strijkers,shashi}  
In FM/SC/FM double junction systems, the coherence length in SC is extracted 
by measuring the magnetoresistance (MR).\cite{taro2,gu}  
In a system consisting of SC with two ferromagnetic leads FM1 and FM2 (see Fig. \ref{fig:Fig_Geometry}), 
there is a novel quantum phenomenon called the crossed Andreev reflection:
\cite{byers,deutscher1,deutscher2,falci,melin1,melin2,melin3,melin4,apinyan,stefanakis,lambert,yu}  
When an electron with energy below the superconducting gap in FM1 is injected into SC, the electron 
captures an electron in FM2 to form a Cooper pair in SC.  As a result, a hole is created in FM2.  
Deutscher and Feinberg \cite{deutscher1} have discussed the crossed Andreev reflection and MR 
by using the theory by Blonder, Tinkham, and Klapwijk (BTK).\cite{BTK}  
They argued that the crossed Andreev reflection should occur when the distance between FM1 and FM2 is of the 
order of or less than the size of the Cooper pairs (the coherence length), and calculated the probability of 
the crossed Andreev 
reflection in the case that both ferromagnetic leads are half metals and the spatial separation of FM1 and FM2 
is neglected (one dimensional model), i.e., the effect of the distance between two ferromagnetic leads on 
the crossed Andreev reflection is not incorporated.  Subsequently, Falci {\it et al.} \cite{falci} 
have discussed the crossed Andreev reflection and the elastic cotunneling in the tunneling limit by using the 
lowest order perturbation of the tunneling Hamiltonian.  However, to elucidate the effect of the crossed Andreev 
reflection on the spin transport more precisely, it is important to explore how the crossed Andreev reflection 
depends on the distance between two ferromagnetic leads as well as on the exchange field of FM1 and FM2, 
for arbitrary transparency of the interface from the metallic limit to the tunneling limit.  

In the present paper, we present a theory of the crossed Andreev reflection in structures consisting of SC 
with two ferromagnetic leads.  By extending the BTK theory to this system, we derive an expression of the 
electric current and calculate the current and MR originated from the crossed Andreev reflection.  
The dependence of the current and MR on the distance ($L$) between FM1 and FM2 is examined.  
It is shown that the dependence of the crossed Andreev reflection on the distance $L$ comes from 
the interference between the wave functions in FM1 and FM2, and the probability decreases rapidly 
as $(k_F L)^{-3}$ with increasing $k_F L$, but not the coherence length of SC,\cite{deutscher1} 
where $k_F$ is the Fermi wave number.  
The current and MR are calculated as functions of the exchange field and the height of the interfacial barriers 
in order to clarify the crossed Andreev reflection in the spin transport of the present system.


\section{Model and Formulation}
\begin{figure}
\includegraphics[width=\columnwidth]{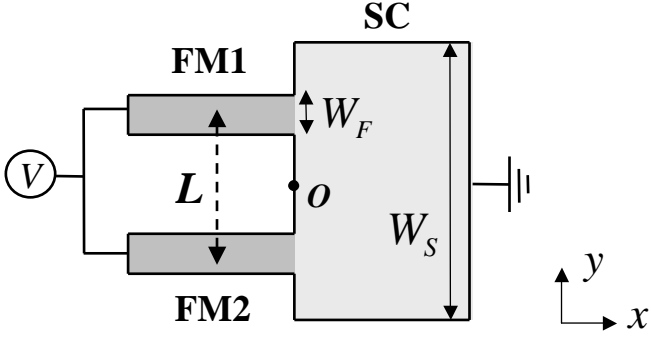}
\caption{Schematic diagram of a superconductor (SC) with two ferromagnetic leads (FM1 and FM2).  
FM1 and FM2 with width $W_F$ are connected to SC with width $W_S$ at $x=0$.  
The distance between FM1 and FM2 is $L$.}
\label{fig:Fig_Geometry}
\end{figure}
We consider a system consisting of a superconductor (SC) with two ferromagnetic leads (FM1 and FM2) as shown 
in Fig. \ref{fig:Fig_Geometry}.  FM1 and FM2 with width $W_F$ are connected to SC with width $W_S$ at 
$x=0$.  The distance between FM1 and FM2 is $L$.  The system we consider is described by the following 
Bogoliubov-de Gennes (BdG) 
equation: \cite{BdG}
\begin{eqnarray}
\left( {\begin{array}{*{20}c}
   {H_0  - \sigma h_{ex} } & \Delta   \\
   {\Delta ^* } & { - \left( {H_0  + \sigma h_{ex} } \right)}  \\
\end{array}} \right)\left( {\begin{array}{*{20}c}
   {f_\sigma  \left( {\bf r} \right)}  \\
   {g_\sigma  \left( {\bf r} \right)}  \\
\end{array}} \right) = E\;\left( {\begin{array}{*{20}c}
   {f_\sigma  \left( {\bf r} \right)}  \\
   {g_\sigma  \left( {\bf r} \right)}  \\
\end{array}} \right) ,\nonumber\\
\end{eqnarray}
where $H_0 \equiv -({\hbar}^2/2m){\nabla}^2-\mu_F$ is the single
particle Hamiltonian measured from the Fermi energy $\mu_F$, 
$E$ is the QP excitation energy, 
and $\sigma=+(-)$ is for the up (down) spin band.  
The exchange field $h_{ex}$ is given by 
\begin{equation}
h_{ex} \left( {\bf r} \right) = \begin{cases}
 \, h_0     & \left( {\, x < 0, \,\, |y-L/2|<W_F/2 \,} \right) ,\\ 
 \, 0       & \left( {\, x > 0  \,}                    \right) ,\\ 
 \, \pm h_0 & \left( {\, x < 0, \,\, |y+L/2|<W_F/2 \,} \right) ,
\end{cases}
\end{equation}
where $+h_{0}$ and $-h_{0}$ represent the exchange fields in FM2 for the parallel
and antiparallel alignments of the magnetizations, respectively.
The superconducting gap is expressed as
\begin{equation}
\Delta \left( {\bf r} \right) = \begin{cases}
 \, \Delta & \left( {\, x > 0, \,\, |y|<W_S/2 \,} \right) ,\\ 
 \, 0      & \left( {x < 0} \right) .
\end{cases}
\end{equation}
We assume that the temperature dependence of the superconducting gap is given by 
$\Delta = \Delta_0 \tanh{\left( 1.74\sqrt{{T_c}/{T}-1}\right)}$,\cite{belzig} 
where $\Delta_0$ is the superconducting gap at $T=0$ and $T_c$ is the superconducting critical temperature.  
In order to capture the effect of the interfacial scattering, 
we employ the following potential at the interfaces, $x=0$:
\begin{equation}
H_{B}\left( {\bf r} \right)
= H \delta \left( {x} \right) \left\{ \theta _1 (y) +  \theta _2 (y) \right\}, 
\end{equation}
where $\delta (x)$ is the delta function and 
$\theta _{1(2)} (y) = \theta \left( {W_F /2 - \left| {y -(+) L/2} \right|} \right)$, 
$\theta (x)$ being the step function.  Throughout this paper, we neglect the impurity scattering in SC and 
the proximity effect near the interfaces.\cite{beenakker,deutscher1,son,perez,halterman1,halterman2}

The solution of the BdG equation in the SC region is given by
\begin{eqnarray}
\begin{array}{l}
\displaystyle \Psi _{ \pm k_{l}^+ } ({\bf r})  = \binom{u_0}{v_0 } e^{ \pm ik_{l}^+ x} \,\Phi_{{\rm SC},l}(y),\\
\displaystyle \Psi _{ \pm k_{l}^- } ({\bf r})  = \binom{v_0}{u_0 } e^{ \pm ik_{l}^- x} \,\Phi_{{\rm SC},l}(y), 
\end{array}
\end{eqnarray}
where ${\bf r}=(x,y)$, and $u_0$ and $v_0$ are the coherence factors, 
\begin{eqnarray}
u_0^2  = 1 - v_0^2  = \frac{1}{2}\left[ {1 + \frac{{\sqrt {E^2 - \Delta ^2 } }}{E}} \right].  
\end{eqnarray}
For $E<\Delta$, $u_0$ and $v_0$ are complex conjugates.  
$\Phi_{{\rm SC},l}(y)$ is the wave function in the $y$ direction, 
\begin{eqnarray}
\Phi_{{\rm SC},l} (y)
= \sqrt {\frac{2}{{W_s }}} \sin \frac{{l\pi }}{{W_s }}\left[ {y + \frac{{W_s }}{2}} \right] , 
\label{wfySC}
\end{eqnarray}
where $l$ is the quantum number which defines the channel.  
The eigenvalue of the $y$ mode for channel $l$ is
\begin{eqnarray}
E_{l} = \frac{\hbar^2}{2m}{\left( \frac{l\pi}{W_S}\right)}^2.
\end{eqnarray}
The $x$ component of the wave number of an electron (hole) like QP, $k_{l}^{+(-)}$, is expressed as
\begin{eqnarray}
k_{l}^{\pm} = \frac{\sqrt{2m}}{\hbar}  \sqrt{\mu _F \pm \sqrt{E^2 - \Delta^2}-E_{l}}.
\end{eqnarray}

In the FM1 (FM2) region, the solutions are given by
\begin{eqnarray}
\begin{array}{l}
\displaystyle \Psi_{\pm p_{\sigma,l}^+}({\bf r})=\binom{1}{0}e^{\pm ip_{\sigma,l}^+ x}\,\Phi_{{\rm FM1 (FM2)},l}(y),\\
\displaystyle \Psi_{\pm p_{\sigma,l}^-}({\bf r})=\binom{0}{1}e^{\pm ip_{\sigma,l}^- x}\,\Phi_{{\rm FM1 (FM2)},l}(y), 
\end{array}
\label{WF-FM}
\end{eqnarray}
where $\Phi _{{\rm FM1 (FM2)},l} (y)$ is the wave function in the $y$ direction 
\begin{eqnarray}
\Phi _{{\rm FM1 (FM2)},l} (y) = \displaystyle \sqrt{\frac{2}{W_F}} 
\sin \frac{{l\pi }}{{W_F }}\left[ {y -(+) \frac{L}{2} + \frac{{W_F }}{2}} \right], \nonumber\\
\label{wfyFM}
\end{eqnarray}
and $p_{\sigma ,l}^{+(-)}$ is the $x$ component of the wave number of an electron (hole)
with $\sigma$ spin;
\begin{eqnarray}
p_{\sigma ,l}^{\pm} = \frac{\sqrt{2m}}{\hbar} \sqrt{\mu _F \pm E \pm \sigma h_{ex} - E_{l}}.
\end{eqnarray}

We consider the scattering of an electron with $\sigma$ spin in channel $n$ injected into SC from FM1.  
There are the following six processes: 
the ordinary Andreev reflection and the normal reflection at the interface of FM1/SC, 
the crossed Andreev reflection, the crossed normal reflection, 
the transmission to SC as an electron like QP, and the one as a hole like QP.  
Therefore, the wave function in each region is expressed as follows: 
In the FM1 region, 

\begin{eqnarray}
\Psi _{\rm FM1} ({\bf r})&=&
\left( {\begin{array}{*{20}c}
   1  \\
   0  \\
\end{array}} \right)e^{ip_{\sigma ,n}^ +  x} \;\Phi _{{\rm FM1},n} (y) \nonumber\\
&+& \sum\limits_{l = 1}^{\infty} \Bigg[ 
a_{\sigma, ln} \left( {\begin{array}{*{20}c}
   0  \\
   1  \\
\end{array}} \right)e^{ip_{\sigma ,l}^ -  x} \nonumber\\
&+&b_{\sigma, ln} \left( {\begin{array}{*{20}c}
   1  \\
   0  \\
\end{array}} \right)e^{ - ip_{\sigma ,l}^ +  x} \Bigg] \;\Phi _{{\rm FM1},l} (y) , 
\label{WF-FM1}
\end{eqnarray}
in the FM2 region, 
\begin{eqnarray}
\Psi _{\rm FM2} ({\bf r})&=&
\sum\limits_{l = 1}^{\infty} \Bigg[ 
c_{\sigma, ln} \left( {\begin{array}{*{20}c}
   0  \\
   1  \\
\end{array}} \right)e^{iq_{\sigma ,l}^ -  x} \nonumber\\
&+& d_{\sigma, ln} \left( {\begin{array}{*{20}c}
   1  \\
   0  \\
\end{array}} \right)e^{ - iq_{\sigma ,l}^ +  x} \Bigg] \;\Phi _{{\rm FM2},l} (y) ,
\label{WF-FM2}
\end{eqnarray}
and in the SC region, 
\begin{eqnarray}
\Psi _{\rm SC} ({\bf r})&=&
\sum\limits_{l = 1}^{\infty} \Bigg[ 
\alpha _{\sigma, ln} \left( {\begin{array}{*{20}c}
   {u_0 }  \\
   {v_0 }  \\
\end{array}} \right)e^{ik_l^ +  x} \nonumber\\
&+& \beta _{\sigma, ln} \left( {\begin{array}{*{20}c}
   {v_0 }  \\
   {u_0 }  \\
\end{array}} \right)e^{ -ik_l^- x} \Bigg] \Phi _{{\rm SC},l} (y) .
\label{WF-SC}
\end{eqnarray}
Here, $p_{\sigma,l}^{\pm}$, $q_{\sigma,l}^{\pm}$, and $k_{l}^{\pm}$
are the wave numbers in FM1, FM2, and SC, respectively.

The boundary conditions at the interfaces ($x=0$) are as follows:
\begin{eqnarray}
\Psi _{\rm FM1} \theta _1 (y) + \Psi _{\rm FM2} \theta _2 (y) = \Psi _{\rm SC} \theta _S (y) , 
\label{boundary1}\\
\displaystyle\frac{{d\Psi _{\rm SC} }}{{dx}} \theta _S (y)
- \frac{d}{{dx}}\left[ {\Psi _{\rm FM1} \theta _1 (y) + \Psi _{\rm FM2} \theta _2 (y)} \right] \nonumber\\
= \displaystyle\frac{2mH}{\hbar^2}\left[ {\Psi_{\rm FM1}\theta_1 (y)+ \Psi _{\rm FM2} \theta_2 (y)} \right] ,
\label{boundary2}
\end{eqnarray}
where $\theta _S (y) = \theta \left( {W_S /2 - \left| y \right|} \right)$.  
From the boundary conditions, the coefficients $a_{\sigma, ln}$, $b_{\sigma, ln}$, 
$c_{\sigma, ln}$, $d_{\sigma, ln}$, $\alpha_{\sigma, ln}$, and $\beta_{\sigma, ln}$ 
are determined (see Appendix) \cite{szafer,furusaki,takagaki,kikuchi}.  
The probabilities of the Andreev reflection $R_{\sigma,mn}^{1,he}$, the normal reflection $R_{\sigma,mn}^{1,ee}$, 
the crossed Andreev reflection $\tilde{R}_{\sigma,mn}^{1,he}$, 
the crossed normal reflection $\tilde{R}_{\sigma,mn}^{1,ee}$, 
the transmission to SC as an electron like QP, $T_{\sigma,mn}^{1,e'e}$, 
and the one as a hole like QP, $T_{\sigma,mn}^{1,h'e}$, are written as, 
\begin{eqnarray}
\begin{array}{l}
R_{\sigma,mn}^{1,he} = \displaystyle\frac{{p_{\sigma,m}^-}}{{p_{\sigma,n}^+}}\left| {a_{\sigma,mn} } \right|^2 ,\\ 
R_{\sigma,mn}^{1,ee} = \displaystyle\frac{{p_{\sigma,m}^+}}{{p_{\sigma,n}^+}}\left| {b_{\sigma,mn} } \right|^2 ,\\ 
\tilde R_{\sigma,mn}^{1,he} 
= \displaystyle\frac{{q_{\sigma ,m}^-  }}{{p_{\sigma,n}^+  }}\left| {c_{\sigma,mn} } \right|^2 ,\\ 
\tilde R_{\sigma,mn}^{1,ee} 
= \displaystyle\frac{{q_{\sigma ,m}^+  }}{{p_{\sigma,n}^+  }}\left| {d_{\sigma,mn} } \right|^2 ,\\ 
T_{\sigma,mn}^{1,e'e} = 
\begin{cases}
\displaystyle\frac{{k_{\sigma ,m}^+}}{{p_{\sigma ,n}^+}} 
\left( {u_0^2-v_0^2} \right)\left| {\alpha_{\sigma,mn}}\right|^2 & , E>\Delta \\
\hspace{1.7cm} 0 & , E<\Delta 
\end{cases} \\
T_{\sigma,mn}^{1,h'e} = 
\begin{cases}
\displaystyle\frac{{k_{\sigma ,m}^-}}{{p_{\sigma ,n}^+}}
\left( {u_0^2-v_0^2} \right)\left| {\beta_{\sigma,mn}}\right|^2 & , E>\Delta \\
\hspace{1.7cm} 0 & , E<\Delta 
\end{cases}
\end{array}
\label{probabilities}
\end{eqnarray}
where the superscript $e'(h')$ and $1$ in Eq. (\ref{probabilities}) indicate 
the electron (hole) like QP in SC and the injection from FM1, respectively.

Let us evaluate the current in FM1.  
When the bias voltage $V$ is applied to the system (see Fig. \ref{fig:Fig_Geometry}), 
the current carried by electrons with $\sigma$ spin in channel $m$ is given by 
\begin{eqnarray}
I_{\sigma,m}^{1,e} = \frac{e}{h}\int_0^\infty 
{\left[ {f_{\sigma,m,\to}^{1,e} \left( E \right) - f_{\sigma,m,\leftarrow}^{1,e} \left( E \right)} \right]} dE, 
\end{eqnarray}
where $h$ is Planck constant, 
and $f_{\sigma,m,\to}^{1,e} \left( E \right)$ is the distribution function of an electron 
with positive group velocity in the $x$ direction and is expressed as 
\begin{eqnarray}
f_{\sigma,m,\to}^{1,e} \left( E \right) = f_0 \left( {E - eV} \right) ,
\end{eqnarray}
where $f_0 \left( E \right)$ is the Fermi distribution function.  
The distribution function of electron with negative group velocity in the $x$ direction 
$f_{\sigma,m,\leftarrow}^{1,e} \left( E \right)$ is given by
\begin{eqnarray}
f_{\sigma,m,\leftarrow}^{1,e} \left( E \right) 
&=& \sum\limits_{l=1}^{\infty} \left[ {R_{\sigma,ml}^{1,ee} f_0 \left( {E - eV} \right)
                                     + R_{\sigma,ml}^{1,eh} f_0 \left( {E + eV} \right)} \right] \nonumber\\
&+& \sum\limits_{l=1}^{\infty} {\left[ {\tilde R_{\sigma,ml}^{2,ee} f_0 \left( {E - eV} \right)
                                      + \tilde R_{\sigma,ml}^{2,eh} f_0 \left( {E + eV} \right)} \right]} \nonumber\\
&+& \sum\limits_{l=1}^{\infty} {\frac{{v_{S ,l}N_{S,l}}}{{v_{F,m}^{\sigma}N_{F,m}^{\sigma}}}
\left[ {T_{\sigma,ml}^{1,ee'} + T_{\sigma,ml}^{1,eh'} } \right]} f_0 \left( E \right) , 
\end{eqnarray}
where $v_{S,l}$ and $v_{F,l}^{\sigma}$ are the group velocity of an electron in channel $l$ in SC and 
the one with $\sigma$ spin in channel $l$ in FM1, respectively, 
$N_{S,l}$ and $N_{F,l}^{\sigma}$ are the density of states in channel $l$ in SC and 
the one of $\sigma$ spin band in channel $l$ in FM1, respectively.
Using the relations, 
\begin{eqnarray}
&R_{\sigma,ml}^{1,ee(eh)} = R_{\sigma,lm}^{1,ee(he)}, \nonumber\\
&\tilde R_{\sigma,ml}^{2,ee(eh)} = \tilde R_{\sigma,lm}^{1,ee(he)}, \nonumber\\
&v_{S ,l}N_{S,l} \,T_{\sigma,ml}^{1,ee'(eh')} = v_{F,m}^{\sigma}N_{F,m}^{\sigma} \,T_{\sigma,lm}^{1,e'e(h'e)}, 
\end{eqnarray}
and the conservation law of the probability, 
\begin{eqnarray}
\sum\limits_{l=1}^\infty 
\Bigg[ \left( {R_{\sigma,lm}^{1,ee} + R_{\sigma,lm}^{1,he} 
      + \tilde R_{\sigma,lm}^{1,ee} + \tilde R_{\sigma,lm}^{1,he} } \right) \nonumber\\
     + \left( {T_{\sigma,lm}^{1,e'e} + T_{\sigma,lm}^{1,h'e} } \right) \Bigg] = 1, 
\label{conservation}
\end{eqnarray}
we obtain
\begin{eqnarray}
&&I_{\sigma,m}^{1,e} = \displaystyle\frac{e}{h} \sum\limits_{l=1}^{\infty} \nonumber\\ 
&\times&\int_0^\infty 
\bigg[ \left( {R_{\sigma,lm}^{1,he} + \tilde{R}_{\sigma,lm}^{1,he} } \right) 
\left[ {f_0 \left( {E} \right) - f_0 \left( {E + eV} \right)} \right] \nonumber\\
&+&\left( {1 - R_{\sigma,lm}^{1,ee} - \tilde{R}_{\sigma,lm}^{1,ee} } \right) 
\left[ {f_0 \left( {E - eV} \right) - f_0 \left( {E} \right)} \right] \bigg] dE.  \nonumber\\
\label{current1}
\end{eqnarray}
The current carried by holes with $\sigma$ spin in channel $m$ in FM1, $I_{\sigma,m}^{1,h}$, 
the currents carried by electrons and holes in FM2, $I_{\sigma,m}^{2,e}$ and $I_{\sigma,m}^{2,h}$, respectively, 
are calculated in the similar way as  
\begin{eqnarray}
&&I_{\sigma,m}^{1,h} = \displaystyle\frac{e}{h} \sum\limits_{l=1}^{\infty} \nonumber\\ 
&\times&\int_0^\infty 
\bigg[ \left( {R_{\sigma,lm}^{1,eh} + \tilde{R}_{\sigma,lm}^{1,eh} } \right) 
\left[ {f_0 \left( {E - eV} \right) - f_0 \left( {E} \right)} \right] \nonumber\\
&+&\left( {1 - R_{\sigma,lm}^{1,hh} - \tilde{R}_{\sigma,lm}^{1,hh} } \right) 
\left[ {f_0 \left( {E} \right) - f_0 \left( {E + eV} \right)} \right] \bigg] dE,  \nonumber\\
\label{current2}\\
&&I_{\sigma,m}^{2,e} = \displaystyle\frac{e}{h} \sum\limits_{l=1}^{\infty} \nonumber\\ 
&\times&\int_0^\infty 
\bigg[ \left( {R_{\sigma,lm}^{2,he} + \tilde{R}_{\sigma,lm}^{2,he} } \right) 
\left[ {f_0 \left( {E} \right) - f_0 \left( {E + eV} \right)} \right] \nonumber\\
&+&\left( {1 - R_{\sigma,lm}^{2,ee} - \tilde{R}_{\sigma,lm}^{2,ee} } \right) 
\left[ {f_0 \left( {E - eV} \right) - f_0 \left( {E} \right)} \right] \bigg] dE,  \nonumber\\
\label{current3}\\
&&I_{\sigma,m}^{2,h} = \displaystyle\frac{e}{h} \sum\limits_{l=1}^{\infty} \nonumber\\ 
&\times&\int_0^\infty 
\bigg[ \left( {R_{\sigma,lm}^{2,eh} + \tilde{R}_{\sigma,lm}^{2,eh} } \right) 
\left[ {f_0 \left( {E - eV} \right) - f_0 \left( {E} \right)} \right] \nonumber\\
&+&\left( {1 - R_{\sigma,lm}^{2,hh} - \tilde{R}_{\sigma,lm}^{2,hh} } \right) 
\left[ {f_0 \left( {E} \right) - f_0 \left( {E + eV} \right)} \right] \bigg] dE.  \nonumber\\
\label{current4}
\end{eqnarray}
By using Eqs. (\ref{current1})-(\ref{current4}), we obtain the total current in the system 
\begin{eqnarray}
I = \sum_{\sigma,m} 
\left[ {I_{\sigma,m}^{1,e} + I_{\sigma,m}^{1,h} + I_{\sigma,m}^{2,e} + I_{\sigma,m}^{2,h} }\right] .  
\label{currenttotal}
\end{eqnarray}
We define the magnetoresistance (MR) as 
\begin{equation}
{\rm MR} \equiv \frac{R_{\rm AP} - R_{\rm P}}{R_{\rm P}} = \frac{I_{\rm P} - I_{\rm AP}}{I_{\rm AP}}, 
\label{def-MR}
\end{equation}
where $R_{\rm P(AP)} = V/I_{\rm P(AP)}$ is the resistance in the parallel (antiparallel) alignment.


\section{Results}

\begin{figure}
\includegraphics[width=\columnwidth]{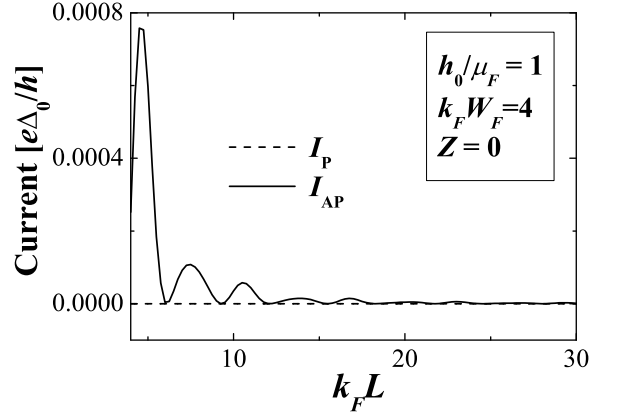}
\caption{The current as a function of $L$.  
FM1 and FM2 are half metals ($h_0 / \mu_F = 1$).  
The solid and dashed lines are for the currents 
in the antiparallel and parallel alignments of the magnetizations, respectively.}
\label{Fig_I-HM}
\end{figure}

In the following calculation, we take the temperature, the applied bias voltage, the width of SC, 
and the superconducting order parameter to be $T/T_c = 0.01$, $eV/\Delta_0 = 0.01$, $W_S = 1000/k_F$, 
and $\mu_F/\Delta_0 = 200$, respectively, where $k_F$ is the Fermi wave number.  
First, we consider the case that FM1 and FM2 are half metals ($h_{0}/\mu_F = 1$) 
and the strength of the interfacial barrier $Z=mH/\hbar^2 k_F=0$.  The width of FM1 and FM2 is taken to be 
$W_F=4/k_{F}$.  In this case, there is only one propagating mode ($l=1$ in Eq. (\ref{WF-FM})).  
We obtain the maximum possible value of MR, i.e., ${\rm MR} = -1$ independently of $L$.  
In order to understand this behavior, we consider the $L$ dependence of 
the currents in the parallel and antiparallel alignments as shown in Fig \ref{Fig_I-HM}.  
When an electron with up spin in FM1 is injected into SC, 
the ordinary Andreev reflection does not occur because electrons with down spin are absent in FM1.  
In the parallel alignment, the crossed Andreev reflection does not occur either 
because there are no electrons with down spin in FM2.  
Therefore, no current flows in the system as shown in Fig \ref{Fig_I-HM}.  
On the other hand, in the antiparallel alignment, while the ordinary Andreev reflection is absent, 
the crossed Andreev reflection occurs because there are electrons with down spin in FM2, which is a member 
of a Cooper pair, for an incident electron with up spin from FM1, 
and therefore finite current flows in the system as shown in Fig \ref{Fig_I-HM}.  
As a result, we find ${\rm MR} = -1$ irrespective of $L$ in the case of half metallic FM1 and FM2.  
The current in the antiparallel alignment decreases oscillating with increasing $L$.  
The behavior of the current is understood as follows.  
From Eqs. (\ref{current1})-(\ref{currenttotal}), 
the current in the antiparallel alignment at low temperatures and low applied bias voltage is expressed as
\begin{eqnarray}
I_{\rm AP} \sim \frac{e^2 V}{h} \left[ \tilde{R}_{\uparrow,11}^{1,he} + \tilde{R}_{\downarrow,11}^{1,eh}
 + \tilde{R}_{\downarrow,11}^{2,he} + \tilde{R}_{\uparrow,11}^{2,eh} \right].  
\label{IAP}
\end{eqnarray}
It is shown from Eq. (\ref{condition1})-(\ref{phase}) in the Appendix 
that the $L$ dependence of the probability of the crossed Andreev reflection $\tilde{R}_{\uparrow,11}^{1,he}$ 
is originated from the interference term between the wave functions of FM1 and FM2, and 
\begin{eqnarray}
\tilde{R}_{\uparrow,11}^{1,he} \propto \left[{1 - \sin{(2k_F L + \phi)}}\right] {(k_F L)}^{-3} \exp{(-L/\xi )}, 
\label{Rdependence}
\end{eqnarray}
where $\xi = \xi_{GL}({\pi\Delta}/{2\sqrt{\Delta^2-E^2}})$, $\xi_{GL} = \hbar v_F/\pi\Delta$ being the 
Ginzburg-Landau (GL) coherence length, $v_F$ is the Fermi velocity, and 
$\phi$ is a phase defined as Eq. (\ref{phase}).  
The probabilities $\tilde{R}_{\downarrow,11}^{1,eh}$, $\tilde{R}_{\downarrow,11}^{2,he}$, 
and $\tilde{R}_{\uparrow,11}^{2,eh}$ show the same $L$ dependence as $\tilde{R}_{\uparrow,11}^{1,he}$ 
in Eq. (\ref{Rdependence}), and therefore the current in the 
antiparallel alignment (\ref{IAP}) decreases rapidly with a rate of $(k_F L)^{-3}$ oscillating with period of 
$\pi$ with increasing $k_F L$.  
Note that the $k_F L$ dependence of the probabilities is dominated by the term $(k_F L)^{-3}$, 
not the exponential term $\exp{(-L/\xi )}$,\cite{deutscher1} since $k_F \xi \gg 1$.\cite{tinkham,hsiang}

\begin{figure}
\includegraphics[width=\columnwidth]{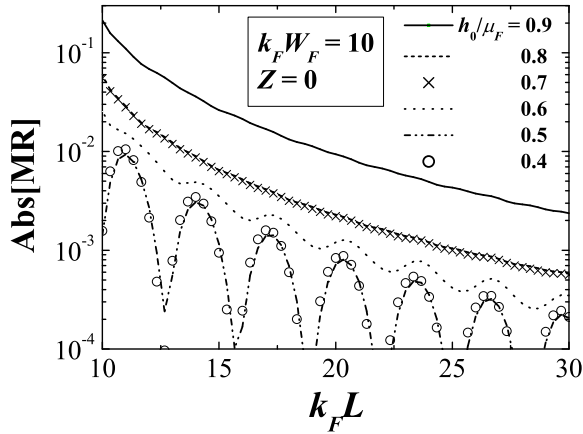}
\caption{The absolute value of MR as a function of $L$ in the case that 
the exchange field $h_0 / \mu_F$ are 0.4, 0.5, 0.6, 0.7, 0.8, and 0.9.}
\label{Fig_MR-L}
\end{figure}

\begin{figure}
\includegraphics[width=\columnwidth]{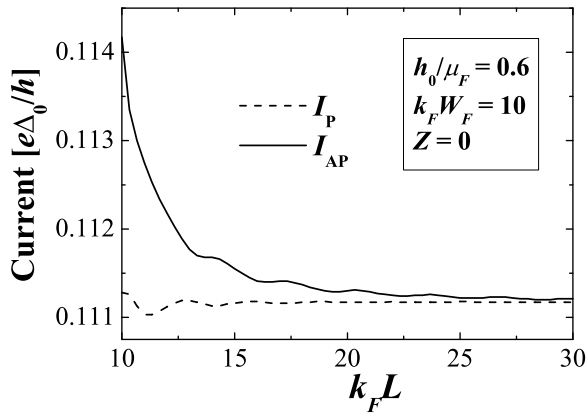}
\caption{The current as a function of $L$ in the case of $h_0 / \mu_F = 0.6$.  
The solid and dashed lines are for the currents in the antiparallel and parallel alignments, respectively.}
\label{Fig_I-L-p06}
\end{figure}

\begin{figure}
\includegraphics[width=\columnwidth]{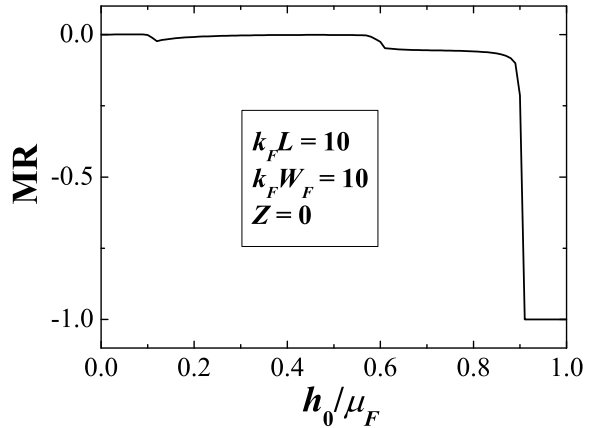}
\caption{MR as a function of $h_0$ for $L = 10/k_F$.}
\label{Fig_MR-pol}
\end{figure}

\begin{figure}
\includegraphics[width=\columnwidth]{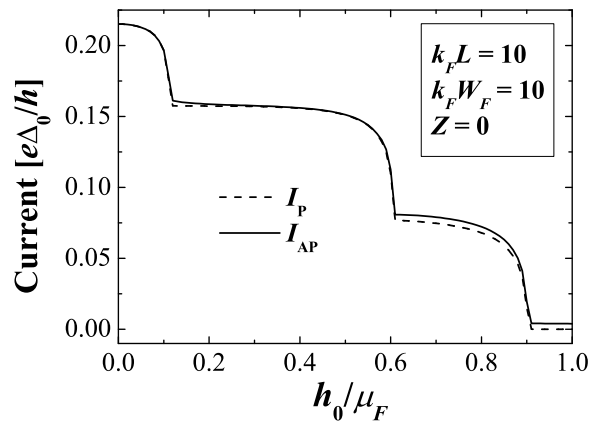}
\caption{The current as a function of $h_0$ for $L = 10/k_F$.  
The solid and dashed lines are for the current in the antiparallel and parallel alignments, respectively.}
\label{Fig_I-pol}
\end{figure}

\begin{figure}
\includegraphics[width=\columnwidth]{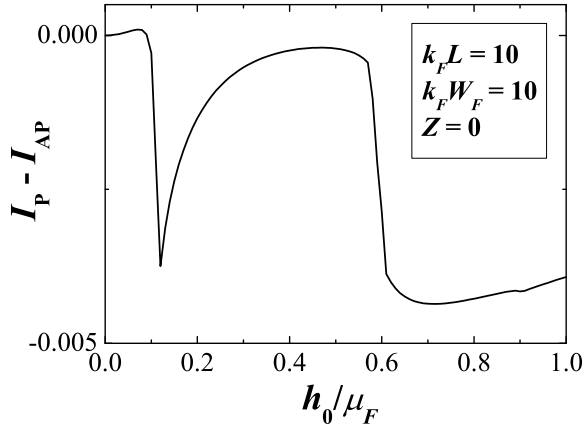}
\caption{The difference between the currents in the parallel and antiparallel alignments as a function of $h_0$.}
\label{Fig_difference-pol}
\end{figure}

We next consider the $L$ dependence of MR for several values of the exchange field in the case that 
$W_F=10/k_{F}$, and $Z=0$ (Fig. \ref{Fig_MR-L}).  
In this case, there are several propagating modes in FM1 and FM2.  
The magnitude of MR decreases with increasing $L$ for each value of the exchange field.  
This behavior of MR is understood by considering the $L$ dependence of the current in the parallel and 
antiparallel alignments.  
As shown in Fig. \ref{Fig_I-L-p06}, in the case that $h_0 = 0.6 \mu_F$, the finite current in the parallel 
alignment flows because the ordinary Andreev reflection occurs, and is almost independent of $L$.  On the other 
hand, the current in the antiparallel alignment decreases with increasing $L$ since the contribution of the 
crossed Andreev reflection process to the current decreases with increasing $L$, 
and therefore the magnitude of MR decreases with increasing $L$.  
In this case, the oscillation of the current in the antiparallel alignment is suppressed 
because electrons and holes in the several propagating modes $l$ in Eq. (\ref{WF-FM}) contribute to the current 
and wash out the oscillation.  
The reason why MR for $h_0 = 0.8 \,(0.5)\, \mu_F$ are almost equal to MR for $h_0 = 0.7 \,(0.4)\, \mu_F$ is as 
follows.  In Fig. \ref{Fig_MR-pol}, the $h_0$ dependence of MR is plotted.  
We find three drops in MR at $h_0 \sim 0.12\, \mu_F$, $0.62\, \mu_F$, and $0.92\, \mu_F$.  
MR for $h_0 = 0.8 \,(0.5)\, \mu_F$ and $h_0 = 0.7 \,(0.4)\, \mu_F$ are in the same plateau.  
This plateau structure is understood by considering 
the denominator $I_{\rm AP}$ and the numerator $I_{\rm P}-I_{\rm AP}$ in Eq. (\ref{def-MR}) separately.  
As shown in Fig. \ref{Fig_I-pol}, $I_{\rm AP}$ is mainly given by the ordinary Andreev reflection, and 
decreases with increasing $h_0$ because the number of the channels for the minority spin decreases by one when 
passing across $h_0 \sim 0.12\, \mu_F$, $0.62\, \mu_F$, and $0.92\, \mu_F$.  Especially, in the range of 
$h_0/\mu_F= 0.92 \sim 1$, there is no open channel for minority spin and the ordinary Andreev reflection is 
completely prohibited.  Therefore, we find ${\rm MR}=-1$ (see Fig. \ref{Fig_MR-pol}).  
Fig. \ref{Fig_difference-pol} shows the $h_0$ dependence of $I_{\rm P}-I_{\rm AP}$, 
which is mainly due to the crossed Andreev reflection.  
The magnitude of $I_{\rm P}-I_{\rm AP}$ is much smaller than that of $I_{\rm AP}$, 
and therefore MR shows the plateau structure as shown in Fig. \ref{Fig_MR-pol}, 
and MR for $h_0 = 0.8 \,(0.5)\, \mu_F$ are almost equal to MR for $h_0 = 0.7 \,(0.4)\, \mu_F$.

\begin{figure}
\includegraphics[width=\columnwidth]{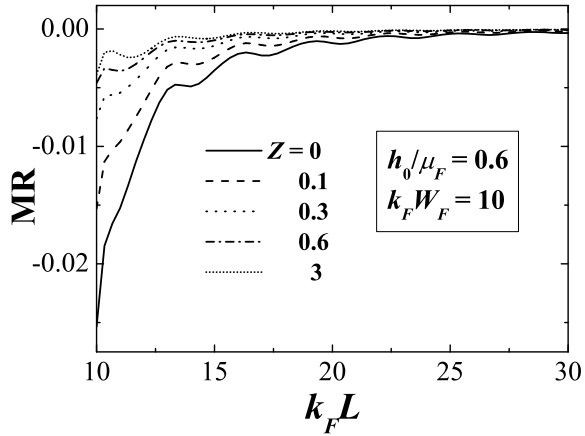}
\caption{MR as a function of $L$ for various values of the interfacial barrier parameter $Z$ 
and $h_0 / \mu_F = 0.6$.}
\label{Fig_MR-L-Z}
\end{figure}

\begin{figure}
\includegraphics[width=\columnwidth]{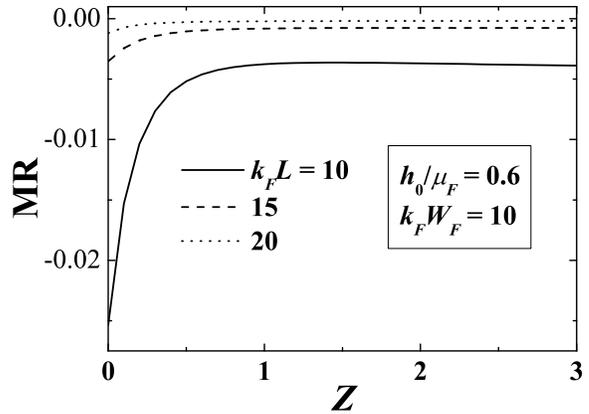}
\caption{MR as a function of the height of the interfacial barriers $Z$ for $h_0 / \mu_F = 0.6$.  
The solid, dashed, and dotted lines represent the case of $k_F L = 10, 15,$ and $20$, respectively.}
\label{Fig_MR-Z}
\end{figure}

Finally, we investigate the effect of the interfacial barriers on the transport in this system.  
Figure \ref{Fig_MR-L-Z} shows the $L$ dependence of MR for $h_0 = 0.6 \mu_F$ 
and several values of interfacial barrier parameter $Z$.  
As seen in Fig. \ref{Fig_MR-L-Z}, MR approaches zero with increasing $L$ 
and shows strong dependence on the height of the interfacial barrier $Z$.  
The decrease of MR with increasing $L$ is explained by the same way as
in the case of no interfacial barriers (Fig. \ref{Fig_MR-L}).  
To investigate the $Z$ dependence of MR in detail, 
we calculate the $Z$ dependence of MR for $k_F L= 10, 15,$ and $20$ as shown in Fig. \ref{Fig_MR-Z}.  
The magnitude of MR decreases with increasing $Z$ in the range of $Z\lesssim 0.5$ and 
is almost constant for $L$ in the range of $Z\gtrsim 0.5$.  
This dependence is understood as follows.  
MR consists of the denominator $I_{\rm AP}$ and the numerator $I_{\rm P}-I_{\rm AP}$, which mainly come from 
the process of the ordinary Andreev reflection and the crossed Andreev reflection, respectively.  
The crossed Andreev reflection is more sensitive to the scattering at the interfacial barriers than 
the ordinary Andreev reflection, and therefore the value of $I_{\rm P}-I_{\rm AP}$ decreases more rapidly 
than that of $I_{\rm AP}$ in the range of $Z\lesssim 0.5$, and therefore the magnitude of MR decreases with 
increasing $Z$ for $k_F L=10, 15,$ and $20$ as shown in Fig. \ref{Fig_MR-Z}.

Although the impurity scattering in SC and the proximity effect are neglected in our theory, these 
assumptions are justified as follows.  First, as shown in the present calculations, the crossed Andreev 
reflection process occurs on the scale which is less than several nanometers for 
$k_F \sim 1\,{\rm \r{A}^{-1}}$.\cite{ashcroft}  This scale is much smaller than the mean free path of 
SC,\cite{hsiang} and therefore the effect of the impurity scattering in SC on the crossed Andreev reflection 
is neglected.  Second, in the present paper, we consider the case that the area of the contacts of FM1 and FM2 
with SC are several nanometers and thus the proximity effect can be neglected.
\cite{beenakker,deutscher1,son,perez}

\section{Conclusion}
We present a theory of the crossed Andreev reflection in structures consisting of a superconductor with two 
ferromagnetic leads.  By extending the BTK theory to this system, we calculate 
the current and the magnetoresistance due to the crossed Andreev 
reflection.  It is shown that the dependence of the crossed Andreev reflection on the distance between two 
ferromagnetic leads, $L$, is given by the interference between the wave functions in 
ferromagnetic leads.  The probability of the crossed Andreev reflection follows $(k_F L)^{-3}$, 
where $k_F$ is the Fermi wave number, and therefore the magnetoresistance due to the crossed Andreev reflection 
strongly decreases with increasing $k_F L$ except for the case of half metallic ferromagnets.  
It is also presented that the dependences of the magnetoresistance on the exchange field show the plateau 
structure and the magnitude of the magnetoresistance rapidly decreases with increasing the height of the 
interfacial barriers.  These dependences are explained by considering the relation between the probabilities of 
the ordinary Andreev reflection and the crossed Andreev reflection.  
\acknowledgements
This work was supported by NAREGI Nanoscience Project, Ministry of
Education, Culture, Sports, Science and Technology (MEXT) of Japan, 
and by a Grant-in-Aid from MEXT and CREST of Japan.
\appendix
\section{boundary conditions}
The coefficients $a_{\sigma, ln}$, $b_{\sigma, ln}$, $c_{\sigma, ln}$, $d_{\sigma, ln}$, 
$\alpha_{\sigma, ln}$, and $\beta_{\sigma, ln}$ in Eqs. (\ref{WF-FM1}), (\ref{WF-FM2}), and (\ref{WF-SC}) 
are determined from the boundary conditions (\ref{boundary1}) and (\ref{boundary2}) as follows. 
\cite{szafer,furusaki,takagaki,kikuchi}  
Substituting the wave functions (\ref{WF-FM1}), (\ref{WF-FM2}), and (\ref{WF-SC})
for the boundary conditions (\ref{boundary1}) and (\ref{boundary2}), we obtain
\begin{eqnarray}
&&\Bigg\{ \left( 
{\begin{array}{*{20}c}
   1  \\
   0  \\
\end{array}} \right)\Phi _{{\rm FM1},n} (y) \nonumber\\
&&+ \sum\limits_{l = 1}^\infty {\left[ {a_{\sigma ,ln} \left( 
{\begin{array}{*{20}c}
   0  \\
   1  \\
\end{array}} \right) + b_{\sigma ,ln} \left( 
{\begin{array}{*{20}c}
   1  \\
   0  \\
\end{array}} \right)} \right]} \;\Phi _{{\rm FM1},l} (y) \Bigg\}\;\theta _1 \left( y \right) \nonumber\\ 
&&+ \sum\limits_{l = 1}^\infty  {\left[ {c_{\sigma ,ln} \left( {\begin{array}{*{20}c}
   0  \\
   1  \\
\end{array}} \right) + d_{\sigma ,ln} \left( {\begin{array}{*{20}c}
   1  \\
   0  \\
\end{array}} \right)} \right]} \;\Phi _{{\rm FM2,}l} (y)\;\theta _2 \left( y \right) \nonumber\\
&&= \sum\limits_{l = 1}^\infty  {\left[ {\alpha _{\sigma ,ln} \left( {\begin{array}{*{20}c}
   {u_0 }  \\
   {v_0 }  \\
\end{array}} \right) + \beta _{\sigma ,ln} \left( {\begin{array}{*{20}c}
   {v_0 }  \\
   {u_0 }  \\
\end{array}} \right)} \right]\;\Phi _{{\rm SC},l} (y)} \;\theta _S \left( y \right) , \nonumber\\
\label{condition1}
\end{eqnarray}
and 
\begin{eqnarray}
&&{\kern 1pt} {\kern 1pt} \Bigg\{ \left( p_{\sigma ,n}^ + -i\frac{2mH}{\hbar^2}\right) 
\left( {\begin{array}{*{20}c}
   1  \\
   0  \\
\end{array}} \right)\Phi _{{\rm FM1},n} (y) \nonumber\\
&&+ \sum\limits_{l = 1}^\infty  
\Bigg[ a_{\sigma ,ln} \left( p_{\sigma ,l}^ - -i\frac{2mH}{\hbar^2}\right)
\left( {\begin{array}{*{20}c}
   0  \\
   1  \\
\end{array}} \right) \nonumber\\
&&- b_{\sigma ,ln} \left( p_{\sigma ,l}^ + +i\frac{2mH}{\hbar^2}\right) 
\left( {\begin{array}{*{20}c}
   1  \\
   0  \\
\end{array}} \right) \Bigg] \;\Phi _{{\rm FM1},l} (y) \Bigg\}\;\theta _1 \left( y \right) \nonumber\\ 
&&+ \sum\limits_{l = 1}^\infty  
\Bigg[ c_{\sigma ,ln} \left( q_{\sigma ,l}^ - -i\frac{2mH}{\hbar^2}\right) 
\left( {\begin{array}{*{20}c}
   0  \\
   1  \\
\end{array}} \right) \nonumber\\
&&- d_{\sigma ,ln} \left( q_{\sigma ,l}^ + +i\frac{2mH}{\hbar^2}\right) 
\left( {\begin{array}{*{20}c}
   1  \\
   0  \\
\end{array}} \right) \Bigg] \;\Phi _{{\rm FM2},l} (y) \;\theta _2 \left( y \right) \nonumber\\ 
&&= \sum\limits_{l = 1}^\infty  {\left[ {\alpha _{\sigma ,ln} k_l^ +  \left( {\begin{array}{*{20}c}
   {u_0 }  \\
   {v_0 }  \\
\end{array}} \right) - \beta _{\sigma ,ln} k_l^ -  \left( {\begin{array}{*{20}c}
   {v_0 }  \\
   {u_0 }  \\
\end{array}} \right)} \right]\;\Phi _{{\rm SC},l} (y)} \;\theta _S \left( y \right) .  \nonumber\\
\label{condition2}
\end{eqnarray}
First, by multiplying the both sides of Eq. (\ref{condition1}) by $\Phi_{{\rm SC},m}(y)$ and 
integrating them with respect to $y$, we obtain 
\begin{eqnarray}
\left( {\begin{array}{*{20}c}
   1  \\
   0  \\
\end{array}} \right)\Lambda _{1,nm} 
&+& \sum\limits_{l = 1}^\infty  {\left[ {a_{\sigma ,ln} \left( {\begin{array}{*{20}c}
   0  \\
   1  \\
\end{array}} \right) + b_{\sigma ,ln} \left( {\begin{array}{*{20}c}
   1  \\
   0  \\
\end{array}} \right)} \right]} \;\Lambda _{1,lm} \nonumber\\
&+& \sum\limits_{l = 1}^\infty  {\left[ {c_{\sigma ,ln} \left( {\begin{array}{*{20}c}
   0  \\
   1  \\
\end{array}} \right) + d_{\sigma ,ln} \left( {\begin{array}{*{20}c}
   1  \\
   0  \\
\end{array}} \right)} \right]} \;\Lambda _{2,lm} \nonumber\\
&=& \alpha _{\sigma ,mn} \left( {\begin{array}{*{20}c}
   {u_0 }  \\
   {v_0 }  \\
\end{array}} \right) + \beta _{\sigma ,mn} \left( {\begin{array}{*{20}c}
   {v_0 }  \\
   {u_0 }  \\
\end{array}} \right) , 
\label{newbou1}
\end{eqnarray}
where $\Lambda_{1(2),lm}(L)$ is the overlap integral between the wave functions in FM1(FM2) and SC, 
and is given by 
\begin{eqnarray}
\Lambda_{1,lm}(L) = 
\displaystyle \int_{(L-W_F)/2}^{(L+W_F)/2} {\Phi_{{\rm FM1},l}(y) \Phi_{{\rm SC},m}(y)} dy, 
\label{lambda1}\\
\Lambda_{2,lm}(L) = 
\displaystyle \int_{(-L-W_F)/2}^{(-L+W_F)/2} {\Phi_{{\rm FM2},l}(y) \Phi_{{\rm SC},m}(y)} dy.  
\label{lambda2}
\end{eqnarray}
Second, by multiplying the both sides of Eq. (\ref{condition2}) by $\Phi_{{\rm FM1},m}(y)$ and 
integrating them with respect to $y$, we obtain
\begin{eqnarray}
&&\left( p_{\sigma ,m}^ + -i\frac{2mH}{\hbar^2} \right) 
\left( {\begin{array}{*{20}c}
   1  \\
   0  \\
\end{array}} \right)\delta _{mn}  \nonumber\\
&&+ a_{\sigma ,mn} \left( p_{\sigma ,m}^ - -i\frac{2mH}{\hbar^2} \right) 
\left( {\begin{array}{*{20}c}
   0  \\
   1  \\
\end{array}} \right) \nonumber\\
&&- b_{\sigma ,mn} \left( p_{\sigma ,m}^ + +i\frac{2mH}{\hbar^2} \right)
\left( {\begin{array}{*{20}c}
   1  \\
   0  \\
\end{array}} \right) \nonumber\\
&&= \sum\limits_{l = 1}^\infty  {\left[ {\alpha _{\sigma ,ln} k_l^ +  \left( {\begin{array}{*{20}c}
   {u_0 }  \\
   {v_0 }  \\
\end{array}} \right) - \beta _{\sigma ,ln} k_l^ -  \left( {\begin{array}{*{20}c}
   {v_0 }  \\
   {u_0 }  \\
\end{array}} \right)} \right]\;\Lambda _{1,ml} } ,  \nonumber\\
\label{newbou2}
\end{eqnarray}
where $\delta _{mn}$ is a Kronecker delta defined as 
\begin{eqnarray}
\delta _{mn}  = \left\{ \begin{array}{l}
 1{\kern 1pt} \;\;{\kern 1pt} (m = n) \\ 
 0\;\;\,(m \ne n) \\ 
 \end{array} \right. .  
\end{eqnarray}
Third, by multiplying the both sides of Eq. (\ref{condition2}) by $\Phi_{{\rm FM2},m}(y)$ and 
integrating them with respect to $y$, we obtain 
\begin{eqnarray}
&&c_{\sigma ,mn} \left( q_{\sigma ,m}^ - -i\frac{2mH}{\hbar^2} \right) 
\left( {\begin{array}{*{20}c}
   0  \\
   1  \\
\end{array}} \right) \nonumber\\
&&- d_{\sigma ,mn} \left( q_{\sigma ,m}^ + +i\frac{2mH}{\hbar^2} \right)  \left( {\begin{array}{*{20}c}
   1  \\
   0  \\
\end{array}} \right) \nonumber\\
&&= \sum\limits_{l = 1}^\infty  {\left[ {\alpha _{\sigma ,ln} k_l^ +  \left( {\begin{array}{*{20}c}
   {u_0 }  \\
   {v_0 }  \\
\end{array}} \right) - \beta _{\sigma ,ln} k_l^ -  \left( {\begin{array}{*{20}c}
   {v_0 }  \\
   {u_0 }  \\
\end{array}} \right)} \right]\;\Lambda _{2,ml} } .  \nonumber\\
\label{newbou3}
\end{eqnarray}

By solving the Eqs. (\ref{newbou1}), (\ref{newbou2}), and (\ref{newbou3}), 
the coefficients $a_{\sigma, ln}$, $b_{\sigma, ln}$, $c_{\sigma, ln}$, $d_{\sigma, ln}$, 
$\alpha_{\sigma, ln}$, and $\beta_{\sigma, ln}$ are determined.  In the numerical calculation, 
we truncate the number of the channels in the ferromagnetic leads (FM1 and FM2) and SC 
by the cutoff constants $M_F$ and $M_S$, respectively.  
The $M_F$ and $M_S$ are taken to be large enough to make the calculation results converge.  

Especially, in the half metallic ($h_0 /\mu_F = 1$) FM1 and FM2 with width $W_F=4/k_{F}$, 
there is only one propagating mode $l=1$ in Eq. (\ref{WF-FM}).  In this case, we can neglect the evanescent 
mode $l \geq 2$ and take the cutoff constant in FM1 and FM2, to be $M_F = 1$.  
In the case of no interfacial barriers ($Z=0$), at low energy region ($E \sim 0$), 
the coefficient of the crossed Andreev reflection part in the wave function of FM2, 
$c_{\uparrow ,11}$, is written as 
\begin{eqnarray}
c_{\uparrow ,11} \sim \frac{C_1 \,\Gamma^+ + C_2 \,\Gamma^-}{C_3}, 
\label{c}
\end{eqnarray}
where 
\begin{eqnarray}
C_1 &=& ip^+ (p^- + \Omega_1^-)(q^+ - \Omega_2^-) ,\label{c1}\\
C_2 &=& ip^+ (p^- - \Omega_1^+)(q^+ + \Omega_2^+) ,\label{c2}\\
C_3 &=& \left[ p^+ p^- + \Omega_1^+ \Omega_1^- - (p^+ - p^-)(\Omega_1^+ - \Omega_1^-)/2 \right] \nonumber\\
&\times&\left[ q^+ q^- + \Omega_2^+ \Omega_2^- - (q^+ - q^-)(\Omega_2^+ - \Omega_2^-)/2 \right], \nonumber\\
\label{c3}
\end{eqnarray}
where $p^{\pm} = p_{\uparrow,1}^{\pm}$, $q^{\pm} = q_{\uparrow,1}^{\pm} $, and 
$\Gamma^{\pm}$ is the interference term between the wave functions of FM1 and FM2 through SC, 
which strongly depends on $L$, given by 
\begin{eqnarray}
\Gamma^{\pm} (L) &=& \sum_{m}^{\infty} {k_{m}^{\pm} \Lambda_{1,1m}(L) \Lambda_{2,1m}}(L) \nonumber\\
&\sim& \frac{k_F\sqrt{2k_F W_F}}{4} {(k_F L)}^{-3/2} \exp{\left( -L/2\xi \right)} \nonumber\\
&&\times \exp{\left( \pm\,i\left( k_F L - \frac{3\pi}{4}\right) \right)} ,
\label{gamma}
\end{eqnarray}
and 
\begin{eqnarray}
\Omega_{1(2)}^{\pm} = \sum_{m}^{\infty} {k_{m}^{\pm} \Lambda_{1(2),1m}^2}.  
\label{omega}
\end{eqnarray}
Substituting Eq. (\ref{c}) for Eq. (\ref{probabilities}), we obtain 
\begin{eqnarray}
&&\tilde{R}_{\uparrow,11}^{1,he} = k_F^3 W_F \nonumber\\
&& \times\, \frac{|C_1|^2+|C_2|^2 
+ C_1 C_2^* e^{i\left({2k_F L-\frac{3\pi}{2}}\right)} 
+ C_1^* C_2 e^{-i\left({2k_F L-\frac{3\pi}{2}}\right)}}{8|C_3|^2} \nonumber\\
&&\times \,{(k_F L)}^{-3} \exp{\left( -L/\xi \right)} \nonumber\\
&&\sim \frac{k_F^3 W_F |C_1|^2}{4|C_3|^2}\nonumber\\
&&\times\left[{1 - \sin{(2k_F L + \phi)}}\right] {(k_F L)}^{-3} \exp{\left( -L/\xi \right)}, 
\end{eqnarray}
where $\phi$ is a phase given by
\begin{eqnarray}
C_1 C_2^* = |C_1 C_2^*| e^{i\phi}, 
\label{phase}
\end{eqnarray}
and we use the relation $|C_1|^2 \simeq |C_2|^2 \simeq |C_1 C_2^*|$.  

\end{document}